\documentclass[twocolumn,twoside,slac_two]{revtex4}
\usepackage{graphicx}
\usepackage{fancyhdr}
\begin{document}

\title{Phenomenological Approach to Multiple Particle Production \\
 {\it --- A model to describe (pseudo-)rapidity density distributions and \\
   transverse momentum distributions in a wide energy region ---}}

\author{A. Ohsawa}
\affiliation{Institute for Cosmic Ray Research, University of Tokyo,
    Kashiwa, 277-8582 Japan.}
\author{E.H. Shibuya}
\affiliation{Instituto de Fisica Gleb Wataghin, Universidade Estadual
de Campinas, Campinas, 13083-970 Brasil.}
\author{M. Tamada}
\affiliation{School of Science and Engineering, Kinki University,
Higashi-Osaka, 577-8502 Japan.}

\begin{abstract}
We formulate empirically the rapidity density distribution 
of produced particles in multiple particle production.
The assumed mechanism is that the produced particles are emitted
isotropically from several emitting centers, located on the
rapidity axis. The formula includes five adjustable parameters, 
which are to be determined by the experimental data of (pseudo-)rapidity
density distributions and transverse momentum distributions
at various energies.  It is a distinguished difference of the present 
rapidity density distribution from those of other models 
that the particle production is suppressed strongly in the forward region.
We discuss multiplicity and inelasticity
at high energies, the pseudo-rapidity density distribution
at LHC energy and some speculations, based on the present formulation. 
\end{abstract}

\maketitle

\thispagestyle{fancy}

\def\pt{p_{\scriptscriptstyle T}}
\section{Introduction}

We formulate the rapidity density distribution of produced
particles {\it phenomenologically} and 
{\it analytically} on the basis of simple assumptions.   
It may clarify what kinds of mechanism are necessary essentially 
to describe multiple particle production and provide a model
which can be extrapolated with more confidence\footnote{We have to say
that most of the models of multiple particle production, proposed so far, 
do not reproduce even the basic data of multiple particle production
reasonably, when we examine them in a wide range of energy.
It means that the models include some inadequate points in them.
We cannot put confidence on the properties of multiple particle
production at high energies which are obtained by extrapolating
such models.} into higher energies. 

\section{Assumptions and formulae}

\subsection{Assumed mechanism of multiple particle production}

We assume the following for the mechanism of multiple particle production. \\
(1) Produced particles are emitted isotropically from several emitting
      centers, which are distributed on the rapidity axis.~(Fig.~1) \\
(2) Produced particles are the {\it newly} produced ones excluding the
   surviving particle.  We assume that all produced particles are pions
   (mass $m$) tentatively. (An effect of kaons among the produced particles
   is not large for the (pseudo-)rapidity density 
   distribution.\cite{ohsawa10})

\begin{figure}
\includegraphics[width=65mm]{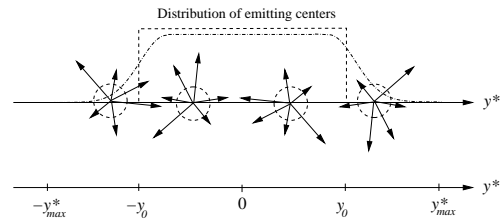}
\caption{Produced particles are emitted isotropically from several
emitting centers, located on the rapidity axis $y^{*}$ in CMS (the
center of mass system). The distribution of emitting centers is
a simple flat one of eq.(2) (the chain line). 
The Fermi distribution of eq.(3) (the chain-dot line) is shown together 
for comparison. $y_{max}^{*} = \ln(\sqrt{s}/M)$ 
($\sqrt{s}$ : the energy in CMS, 
$M$ : nucleon mass) and $y_{0} = y_{max}^{*}-\ln a_{2}$ 
($a_{2}$ : an adjustable parameter).}
\end{figure}

\noindent
(3) The normalized energy distribution of produced particles 
in the rest frames of respective emitting centers is
\begin{equation}
      f(p) dp = {p^{2} \over 2(1+r)} \left[ {1 \over p_{1}^{3}} e^{-p/p_{1}} 
               + {r \over p_{2}^{3}} e^{-p/p_{2}} \right] dp   
\end{equation}
where the values of the parameters $p_{1}$ and $p_{2}$ ($p_{1} < p_{2}$)
are determined
so as to reproduce the $\pt$ (transverse momentum) distributions
at $\sqrt{s}= 63, \; 546$ and $1800$ GeV.
That is, the first and the second term in the energy distribution
correspond to the $\pt$ distributions in the low and high $\pt$ region, 
respectively.
(We showed in Ref.\cite{ohsawa10} that the $\pt$ distribution 
cannot be described
by a single exponential function of the energy distribution.)
The parameter $r$ is the relative contribution of the second term.
(The term $p^{2}$ is necessary to reproduce the $\pt$ distribution 
in the vicinity of $\pt = 0$.) \\
(4) The distribution of the emitting centers in CMS (the center of
mas system) is a ``simple flat'' one;
\begin{equation}
       g(y') dy' = {dy' \over y_{0}}
                \hspace{5mm}(0 \leq y' \leq y_{0})
\end{equation}
where $y_{0} = y_{max}^{*}- \ln a_{2}$ and $y_{max}^{*}= \ln (\sqrt{s}/M)$ 
($\sqrt{s}$ : the CMS energy, $M$ : nucleon mass).\footnote{The quantities 
with an asterisk are those in CMS.}~(see Fig.~1) \par 
The parameter $a_{2}$ describes shrinkage
of the (pseudo-)rapidity density distribution in the forward region, and
the value of it is determined so as to reproduce (pseudo-)rapidity density 
distributions in the region-3\footnote{We divide the rapidity region 
between $y^{*}=0$ and $y^{*}=y_{max}^{*}=\ln (\sqrt{s}/M)$
roughly into three regions, region-1, -2 and -3. The rapidity density
is flat in the region-1, decreasing gradually in the region-2 and
is small, for example, less than a half of the density at $y^{*}=0$,
in the region-3. The pseudo-rapidity region is divided similarly, too.}
at various energies.
Note that the distribution is normalized to one in the forward hemisphere.

We examined another type of the emitting center distribution,
\begin{equation}
     g(y') dy' = {c \over 1 + e^{b(y'-y_{0})}} dy'  
\end{equation}
with $b=2.0$ and $c = b/\ln(1+e^{by_{0}})$,
which will be called ``Fermi distribution''. (see Fig.~1) \\

\subsection{Rapidity density distribution}

The normalized energy-angular distribution of a produced particles 
in the rest frame of an emitting center is 
\[        f(p)dp {1 \over 2} d(\cos \theta)    \]
where the variable $\theta$ is the zenith angle of the produced particle
in the rest frames of respective emitting centers.  
By the variable transformation it turns to
\[       {\pt E \over 2 p^{2}} dy d\pt   \]
where the variable $y$ is the rapidity in the rest frames of respective 
emitting centers. 
Since the rapidity of a produced particle in CMS is $y^{*}=y'+y$,
the rapidity density distribution of charged produced particles
in CMS is
\[    {d^{2} N_{ch} \over dy^{*} d\pt}
           = a_{1} y_{0} \int_{-\infty}^{\infty} {\pt E \over 2p^{2}}
                      f(p) g(y') dy'    \]
\begin{equation}
      = a_{1} \int_{-y_{0}}^{y_{0}} dy'
            {\pt E \over 2(r+1)} \left[ {1 \over p_{1}^{3}} e^{-p/p_{1}}
                    + {r \over p_{2}^{3}} e^{-p/p_{2}} \right]    
\end{equation}
where $E = \mu \cosh(y^{*}-y')$, $p = \sqrt{E^{2}-m^{2}}$ and
$\mu = \sqrt{\pt^{2}+m^{2}}$.
The parameter $a_{1}$ is related to the (pseudo-)rapidity density 
at $y^{*}=0$, and the value of it is determined so as to reproduce
the data at various energies.   Note that the proportional coefficient
is not $a_{1}$ but $a_{1}y_{0}$ in eq.(4).

\subsection{Pseudo-rapidity density distribution and $x$-distribution}

The pseudo-rapidity density distribution and $x$-distribution (defined as
$x^{*}=2p^{*}_{\scriptscriptstyle ||}/\sqrt{s}$, 
$p^{*}_{\scriptscriptstyle ||}$ : longitudinal momentum of produced particle
in CMS) are obtained by variable transformations from eq.(4).
\begin{equation}     
      {d^{2} N_{ch} \over d\eta^{*} d\pt}
             = {\pt(e^{\eta^{*}}+e^{-\eta^{*}}) \over 
                   \sqrt{\pt^{2}(e^{\eta^{*}}+e^{-\eta^{*}})^{2}+4m^{2}}}
                  {d^{2} N_{ch} \over dy{*} d\pt}    
\end{equation}
where
\[     y^{*} = \ln {\pt(e^{\eta^{*}}-e^{-\eta^{*}})
                    + \sqrt{\pt^{2}(e^{\eta^{*}}-e^{-\eta^{*}})^{2}+4\mu^{2}}
                       \over 2\mu}      \]

\begin{equation}   {d^{2} N_{ch} \over dx^{*} d\pt}
               = {1 \over \sqrt{(x^{*})^{2}+(2\mu/\sqrt{s})^{2}}}
                      {d^{2} N_{ch} \over dy^{*} d\pt}            
\end{equation}
where
\[       y^{*} = \ln {\sqrt{(x^{*})^{2}+(2\mu/\sqrt{s})^{2}}+x^{*}
                           \over 2\mu/\sqrt{s}}   \]

\subsection{$\pt$ distribution}

The $\pt$ distribution (at $\theta^{*}=90^{\circ}$)
in terms of the invariant cross section is
\begin{equation}
     \left. E {d^{3} \sigma \over d^{3}p} \right|_{\theta=90^{\circ}}
                = {\sigma_{inel} \over 2\pi \pt}
             \left( {d^{2}N_{ch} \over dy^{*} d\pt} \right)_{y^{*}=0}  
\end{equation} 
The suffix $\theta=90^{\circ}$ will be omitted hereafter.

The local $\pt$ average at the rapidity $y^{*}$ is defined as
\begin{equation}
    < \pt >_{y^{*}} = \left. \int_{0}^{\infty} \pt
                         {d^{2}N_{ch} \over dy^{*}d\pt}  d\pt \right/ 
            \int_{0}^{\infty} {d^{2}N_{ch} \over dy^{*}d\pt}  d\pt    
\end{equation}
The local $\pt$ average at the pseudo-rapidity 
\mbox{$\eta^{*}$, $< \pt >_{\eta^{*}}$},
is defined similarly, which is not the same as eq.(8) in general.

Note that the $\pt$ average, obtained by the experiments, is either
$< \pt >_{y^{*}=0}$ or $< \pt >_{\eta^{*}=0}$ in most of the experiments.

\begin{figure}[t]
\includegraphics[width=65mm]{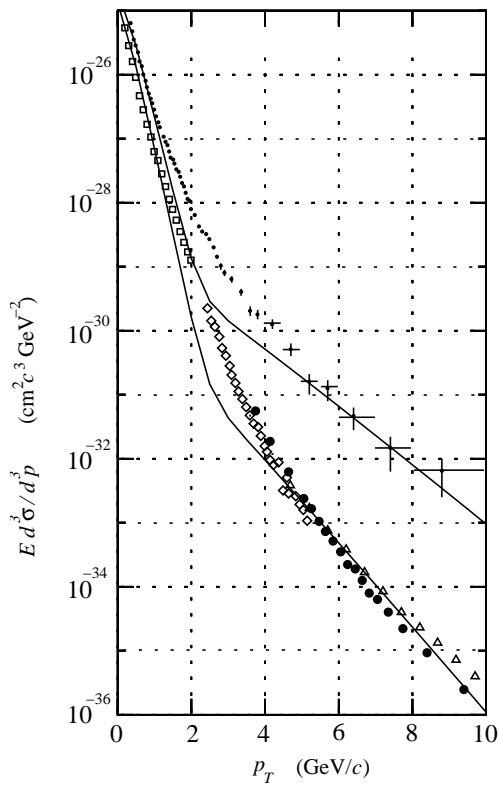}
\caption{The $\pt$ distributions at $\sqrt{s}=63$ GeV (the lower curve) 
and 546 GeV (the upper). Experimental data are compiled 
by UA1 Collaboration.[2] \\
dots~((h$^{+}$+h$^{-}$)/2) : UA1 Collaboration for a half of charged hadrons
at $\sqrt{s}=546$ GeV, squares~((h$^{+}$+h$^{-}$)/2), diamonds~($\pi^{0}$),
triangles~($\pi^{0}$) and circles~((h$^{+}$+h$^{-}$)/2) : ISR data at
$\sqrt{s}=63$ GeV.  Note that value of the curve in the ordinate is
a half of eq.(7) with the values of the parameters in Table~1.}
\end{figure}

\subsection{Multiplicity and inelasticity}

Since the distribution of the emitting centers is normalized in the 
forward hemisphere, charged multiplicity $m_{ch}$ is twice that
of the normalization coefficient in eq.(4),
\begin{equation}     
      m_{ch} = 2 a_{1} y_{0} 
      = 2 a_{1} \left[ \ln \left( {\sqrt{s} \over M} \right)
                                          - \ln a_{2} \right] 
\end{equation}
The total inelasticity in CMS is
\begin{equation}
           K = {3 \over 2} \int_{0}^{1} x^{*} 
                          \left( {dN_{ch} \over dx^{*}} \right) dx^{*}   
\end{equation}

\section{Values of the parameters}

\subsection{Values of the parameters $p_{1}$, $p_{2}$ and $r$}

The values of the parameters in the energy distribution of eq.(1) are
determined by fitting the $\pt$ distribution of eq.(7)
to those of the experiments at $\sqrt{s}=63, \; 546$ and 
$1800$ GeV,\cite{arnison82,abe88} which are shown in Figs.~2 and 3. 

\begin{table*}[t]
\hspace*{4cm}
\caption{Values of the parameters for the $\pt$  distributions}
\begin{tabular}{cccccccc} \hline
$\sqrt{s}$ & $p_{1}$ or $p_{2}$ 
        & \multicolumn{2}{c}{$(dN_{ch}/d\eta^{*})_{\eta^{*}=0}$}
        & $a_{1}$ & $\sigma_{inel}$ & $r$ & Ref. \\
(GeV)    &    (GeV/$c$)  &  cal. & exp. &      &  (mb)  &     &    \\ \hline
         &  $p_{1}=0.154$ &      &      &      &        &     &    \\[-3mm]
63       &        & $0.843a_{1}$ & $1.88 \pm 0.08$ & 2.23 & 36.0 
                         & $2.0 \times 10^{-4}$ & [2]  \\[-3mm]
         &  $p_{2}=0.632$ &      &      &      &        &     &    \\ \hline
         &  $p_{1}=0.1175$ &      &      &      &        &     &    \\[-3mm]
546      &        & $0.865a_{1}$ & $2.79 \pm 0.08$ & 3.23 & 49.0 
                         & $2.0 \times 10^{-3}$ & [2] \\[-3mm]
         &  $p_{2}=0.895$ &      &      &      &        &     &    \\ \hline
         &  $p_{1}=0.206$ &      &      &      &        &     &    \\[-3mm]
1800     &        & $0.875a_{1}$ & $3.95^{*}$          & 4.51 & 56.0 
                         & $2.0 \times 10^{-3}$ & [3] \\[-3mm]
         &  $p_{2}=0.938$ &      &      &      &        &     &    \\ \hline
         &  $p_{1}=0.253$ &      &      &      &        &     &    \\[-3mm]
$1.4 \times 10^{4}$  &  & $-$  & $-$             & 6.54 & 73.0$^{\dagger}$ 
                         & $6.5 \times 10^{-2}$ &     \\[-3mm]
         &  $p_{2}=0.632$ &      &      &      &        &     &    \\ \hline
\multicolumn{8}{l}{$^{*}$ : the value at $\eta^{*}=0.12$, \
      $^{\dagger}$ : Model 2 in Ref.\cite{horandel03}}
\end{tabular}
\end{table*}

Values of the parameters determined are tabulated in Table~1. 
The values of the parameters $p_{1}$ and $p_{2}$ are determined to reproduce
the slopes of the $\pt$ distributions in the $\pt$ regions of
$0.2 - 1.0$ and $5.0 - 10.0$ (GeV/$c$), respectively.  
The value of the parameter $a_{1}$ in Table~1 is determined by equating the 
calculated pseudo-rapidity density at $\eta^{*}=0$
to that of the experiment.\cite{alner87} 
(In the calculation we assume the value of
the parameter $a_{2}=1.0$, which affects slightly the pseudo-rapidity 
density at $\eta^{*}=0$.)
One can see in the figures that the $\pt$ distributions are described well 
in the small and large $\pt$ regions, but not in the middle $\pt$ region.

\begin{figure}[t]
\includegraphics[width=65mm]{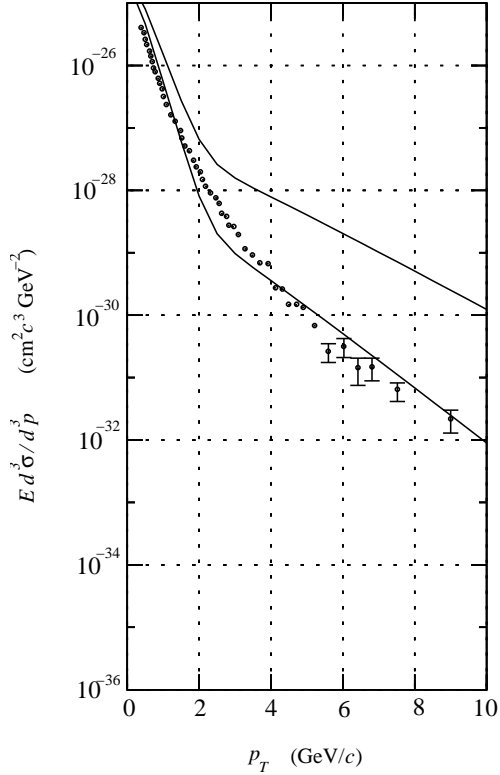}
\caption{The $\pt$ distributions at 1800 GeV (the lower curve) and 
14 TeV (the upper).   Experimental data are for a half of charged hadrons, 
(h$^{+}$+h$^{-}$)/2, from CDF Collaboration.[3]
Note that value of the curve in the ordinate is
a half of eq.(7) with the values of the parameters in Table~1.}
\end{figure}

Fig.~4 shows the energy dependences of the parameters $p_{1}$, $p_{2}$
and $r$ in Table~1.  Assuming the power dependences of the energy, empirical
formulae of the parameters are;
\begin{equation}    
\begin{array}{l}
     p_{1}= 0.0895 (\sqrt{s})^{0.109}  \\
     p_{2} = 0.381 (\sqrt{s})^{0.130} \\
     r = 2.38 \times 10^{-6} (\sqrt{s})^{1.07}     
\end{array}
\end{equation}
where the energy $\sqrt{s}$ is in GeV.

\begin{figure}
\includegraphics[width=65mm]{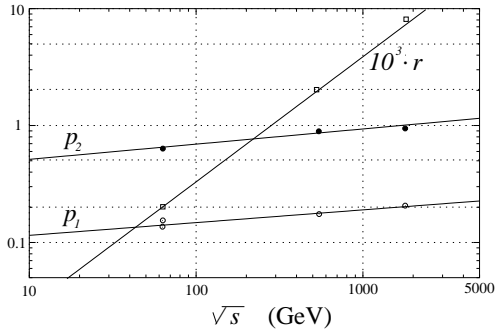}
\caption{Energy dependences of the parameters $p_{1}$ (GeV/$c$) 
(open circles), $p_{2}$ (GeV/$c$) (full circles) and $r$ (squares).
Note that the parameter $r$ is multiplied by a factor $10^{3}$.
The solid lines are eq.(11), the best-fit ones to the data points, 
assuming the power dependences of the energy $\sqrt{s}$.}
\end{figure}

We expected that the parameters $p_{1}$ and $p_{2}$ are energy-independent,
which is not correct.  The value of the parameter $r$ exceeds 1.0
for the energy $\sqrt{s} > 1.80 \times 10^{5}$ GeV, since the exponent
is as large as 1.07.  We can append, however, that the cross section of
mini-jets ($E_{\scriptscriptstyle T} > 5$ GeV, $|\eta^{*}| < 1.5$) increases 
similarly in the energy region $\sqrt{s}= 200 - 900$ GeV.\cite{scott87}
We should note that the exponent may become smaller than 1.07 
at high energies, since the channel of mini-jets has opened just 
in the present energy region.

Fig.~5 shows the $\pt$ average at $\eta^{*}=0$, $< \pt >_{\eta^{*}=0}$, 
and the energy $\sqrt{s}$, based on the energy dependences of the
parameters in eq.(11).  It is of no wonder that experimental data are
consistent with that at $\eta^{*}=0$ (but not at $y^{*}=0$), since
the condition to sample the events refers the pseudo-rapidity
in the data concerned.

\begin{figure}
\includegraphics[width=65mm]{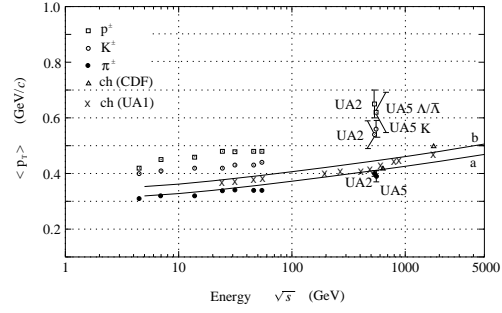}
\caption{The $\pt$ average and the energy $\sqrt{s}$.  
The data are those compiled by UA5 Collaboration[2]
and by UA1 Collaboration[8] and those from CDF Collaboration.[3]
The curves are $< \pt >_{y^{*}=0}$ (denoted as ``a'') and 
$< \pt >_{\eta^{*}=0}$ (``b''), which are defined in 2.4.}
\end{figure}

Fig.~6 shows the local $\pt$ average at the rapidity $y^{*}$
for various values of the parameter $a_{2}$,
together with the data in the region-3. The data are
from UA7 Collaboration at $\sqrt{s}=630$ GeV.\cite{pare90}
The data are described well by the curve of $a_{2}=5.0$.
It is important to note that the rapidity density distribution
in the region-3, obtained by the same collaboration, is
described by the same value of the parameter $a_{2}$
simultaneously.~(see 3.2)

\begin{figure}
\includegraphics[width=65mm]{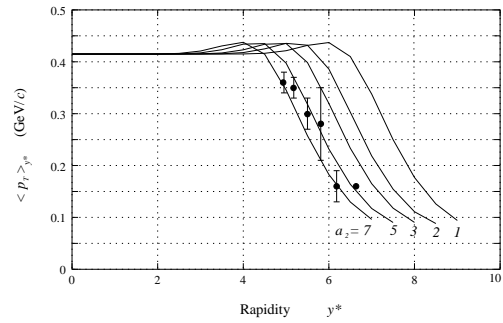}
\caption{The local $\pt$ average of eq.(8) at $\sqrt{s}=630$ GeV.
The curves are for the values of the parameter $a_{2}=1.0, \; 2.0, \; 3.0,
\; 5.0$ and 7.0 (attached to the curves). Data are for $\pi^{0}$'s
by UA7 Collaboration.[6]   The data are described well by the curve
of $a_{2}=5.0$.}
\end{figure}

\subsection{Values of the parameters $a_{1}$ and $a_{2}$}

The values of the parameters $a_{1}$ and $a_{2}$ are determined 
by fitting the (pseudo-)rapidity density distribution, eqs.(4) or (5),
to those of experiments at various energies.
Note that the surviving particle is included among the observed particles
in most of the experimental data.  The density of 
the surviving particle, however, occupies a small part of
the \mbox{(pseudo-)} rapidity density in the region-1 and -2 where
the data exist. (see, for example, Fig.~13)

Sources of the experimental data of the \mbox{(pseudo-)} \\
rapidity density distribution are tabulated in Table~2.  
The data in the table concern only those
that are necessary for the present analysis.  For example, EHS-NA22
Collaboration observed $K^{+}p$ and $\pi^{+}p$ collisions, too.
Necessary comments to respective sets of data are found 
in Ref.\cite{ohsawa10}.

\begin{table*}[t]
\caption{Data of (pseudo-)rapidity density distributions
and the values of the parameters $a_{1}$ and $a_{2}$}
\begin{tabular}{llrcllcccr}\hline
        &      & Energy           &           & Observed  & Observed
       & Observed    & \multicolumn{2}{c}{Parameter}  &      \\
Collab. & Site & $\sqrt{s}$ (GeV) & Collision & particles$^{\dagger}$ & range 
     & quantity    & \multicolumn{1}{c}{$a_{1}$} 
                   & \multicolumn{1}{c}{$a_{2}$} & Ref. \\ \hline 
EHS-NA22 & CERN SPS & 22.4 & $pp$ & $\pi^{-}, \; c^{+}$ 
        & $|\eta^{*}| \leq 6.0$ & $d\sigma/d\eta^{*}$ & 1.8 & $1.5 \pm 0.5$
                 & [9]  \\
Phobos & BNL RHIC & 200 & $pp$ & $c^{\pm}$ & $|\eta^{*}| \leq 5.4$
        & $dN/d\eta^{*}$ & 2.6 & $4 \pm 1$  & [10]  \\
UA5 & CERN SPS & 53 & $\bar{p}p$ & $c^{\pm}$ & $|\eta^{*}| \leq 3.5$
        & $dN/d\eta^{*}$ & 2.1 & $3 \pm 1$  & [4] \\
    &          & 200 & &                    & $|\eta^{*}| \leq 4.6$
    &    &  2.6  & $4 \pm 1$  &    \\ 
    &          & 546 & &                    & $|\eta^{*}| \leq 4.6$
    &    &  3.1  & $8.5 \pm 1.5$  &    \\ 
    &          & 900 & &                    & $|\eta^{*}| \leq 4.6$
    &    & 3.5  &$10 \pm 1$  &    \\ 
UA7 & CERN SPS & 630 & $\bar{p}p$ & $\gamma$ & $y^{*}=5.0-6.6$
       & $d\sigma_{\pi^{0}}/dy^{*}$ & (3.5) & $5 \pm 1$  & [6] \\
P238 & CERN SPS  & 630 & $\bar{p}p$ & $c^{\pm}$ & $\eta^{*}=1.5-5.5$
       & $dN/d\eta^{*}$ & 3.5 & $6 \pm 1$  & [11] \\
CDF & FNAL Tevatron & 630 & $\bar{p}p$ & $c^{\pm}$ & $|\eta^{*}| \leq 3.5$
       & $dN/d\eta^{*}$ & 3.6 & $-$  & [12] \\ 
    &              & 1800 &            &           & 
       &  & 4.4 & $-$  &  \\ \hline
\multicolumn{10}{l}{$^{\dagger}$ The letter $c$ stands for charged particles.}
\end{tabular}
\end{table*}

Figs.~7, 8 and 9 are the examples to show how well the experimental data
are described by the curves of the present formulation.
The curves are calculated for several assumed values of the parameters
$a_{1}$ and $a_{2}$, {\it i.e.} $a_{1}=1.0$ and $a_{2}=1.0, \; 2.0, \;
\cdots$.  Hence, in the figures, the calculated curves are shifted upwards
to fit to the data in the region-1, and then the curve, which fits
best to the data in the region-2 and/or -3, is selected.  This procedure
determines the values of the parameters $a_{1}$ and $a_{2}$, which are
tabulated in Table~2.  (The value of the parameter $a_{2}$ cannot
be determined for the data by CDF Collaboration, since the data
exit only in the region-1.)
We list some comments below to the figures.

\begin{figure}[h]
\includegraphics[width=65mm]{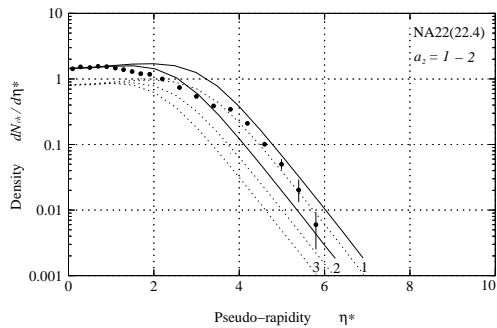}
\caption{Pseudo-rapidity density distribution at $\sqrt{s}=22.4$ GeV.
The data are the sum of $\pi^{-}$ (negative pions) and $c^{+}$ (positive
particles) by EHS-NA22 Collaboration.[9]  Dotted curves are eq.(5)
with the values of the parameters $a_{1}=1.0$ and $a_{2}=1.0, \; 2.0$ 
and 3.0 (attached to the curve).  The solid lines are the best-fit ones
to the experimental data with the values of the parameters $a_{1}=1.8$
and $a_{2}=1.0-2.0$.  Agreement between the curves and the data is poor
in the region-2.}
\end{figure}

\noindent
(1) In Fig.~7 the agreement between the best-fit curves and the data
is good in the region-1 and -3, but is poor in the region-2. 
This situation is the same for almost all sets
of data in Table~2. In Ref.\cite{ohsawa10} we showed 
that the Fermi distribution for the emitting centers, eq.(3), 
brings better agreement 
in the region-2, but that it does not describe the data 
of the local $\pt$ average in the region-3, obtained by UA7 Collaboration.     
One can see, however, in 3.1 that the simple flat distribution
for the emitting centers describe the local $\pt$ average naturally,
which is the reason why we adopt it in the present paper. 

\begin{figure}[h]
\includegraphics[width=65mm]{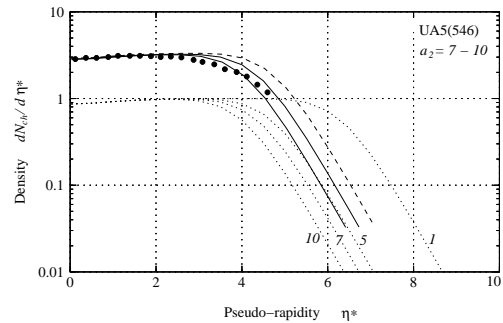}
\caption{Pseudo-rapidity density distribution at $\sqrt{s}=546$ GeV.
Data are by UA5 Collaboration.[4] Dotted curves are eq.(5)
with the values of the parameters $a_{1}=1.0$ and $a_{2}=1.0, \; 5.0,
\; 7.0$ and 10.0 (attached to the curve).  
The solid lines are the best-fit ones
to the experimental data with the values of the parameters $a_{1}=3.1$
and $a_{2}=7.0-10.0$.  The chain line is for the values of the parameters
$a_{1}=3.1$ and $a_{2}=5.0$, which corresponds approximately 
to the assumed case of
the inelasticity $K = 0.5$. (see eq.(12))   Note that the chain curve
{\it does not} describe the data clearly.}  
\end{figure}

\noindent
(2) Fig.~8 shows the pseudo-rapidity density distributions at $\sqrt{s}=546$
GeV, together with the experimental data by UA5 Collaboration.\cite{alner87}
The curves of $a_{1}=3.1$ and $a_{2}=7$ and 10 describe the data
reasonably.

\noindent
(3) In Fig.~9 the value of the parameter $a_{1}=3.5$ for the data by
UA7 Collaboration\cite{pare90} is the assumed one, since data exist only in
the region-3. It is important to note that the rapidity
density distribution is described well by the value 
of the parameter $a_{2}=5.0$ which describes the local
$\pt$ average in Fig.~6.

\begin{figure}[h]
\includegraphics[width=65mm]{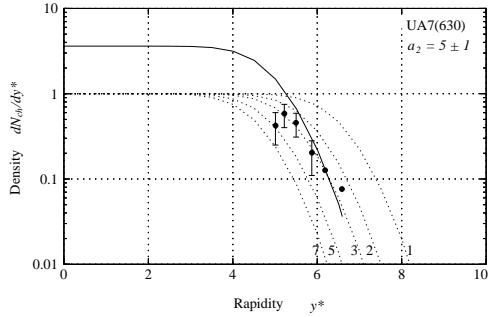}
\caption{Rapidity density distribution  at $\sqrt{s}=630$ GeV.
Data of $\pi^{0}$'s, by UA7 Collaboration[6], are  multiplied 
by a factor 2.0. 
Dotted lines are those for the values of the parameters $a_{1}=1.0$
and $a_{2}=1.0, \; 2.0, \; 3.0, \; 5.0$ and 7.0 (attached to the curves).
The solid line is the best-fit one with the values of the parameters
$a_{1}=3.5$ (assumed) and $a_{2}=5.0$.}  
\end{figure}

Fig.~10 shows the energy dependences of the parameters $a_{1}$ and $a_{2}$
in Table~2.  We list some comments to Fig.~10. \\
(1) Since the (pseudo-)rapidity density distribution is almost flat 
in the region-1, it is easy to fit the data to the calculated curve.
Consequently the values of the parameter $a_{1}$ are determined reliably
to reproduce the data. \\
(2) In the accelerator experiments it is not easy to obtain the data
in the region-2 and -3 due to the experimental conditions. Consequently
the data in the concerned regions are limited or missed in most of
the experiments in Table~2, and the data in the region-2, if they exist,
are often not consistent one another by the experiments 
even at the same incident energy. 
Consequently the values of the parameter
$a_{2}$ are determined with large errors and are distributed widely. \\ 
(3) In order to consider the {\it assumed} case of the inelasticity $K = 0.5$,
we examine the energy dependence of the parameter 
\begin{equation}
   a_{2} = 0.718 (\sqrt{s})^{0.320}
\end{equation} 
As can be seen in Fig.~12, eq.(12) brings the inelasticity $K \simeq 0.5$,
taking the value of the parameter $a_{1}$ as it is, one of eq.(13). 
In Fig.~10 the chain-dot line of eq.(12) 
is almost consistent with points except those at $\sqrt{s}= 546$ and
900 GeV, both of which  are from UA5 Collaboration.
(Fig.~8 shows that the pseudo-rapidity density distribution at $\sqrt{s}=
546$ GeV by UA5 Collaboration is not described by the curve 
of the assumed case of $K = 0.5$ clearly.)
It is not evident, however, that both data are biased 
in the region-2 seriously.
Since it is our strategy in the present paper to formulate 
multiple particle production phenomenologically 
avoiding {\it a priori} assumptions as much as possible, 
we determine the energy dependence of the parameter $a_{2}$
by the least square method including both data. 

\begin{figure}
\includegraphics[width=65mm]{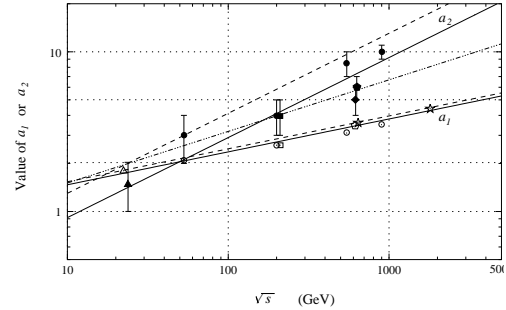}
\caption{The values of the parameters $a_{1}$ (open marks) 
and $a_{2}$ (full marks) in Table~2 and the energy $\sqrt{s}$.
Marks: triangles~(EHS-NA22 Collaboration), circles~(UA5), squares~(Phobos),
pentagons~(P238), diamonds~(UA7), stars~(CDF).
Solid lines are the best-fit ones to the data points, eq.(13) in the text.
Chain lines are those in Ref.[1].
The chain-dot line is eq.(12) in the text,
for which the total inelasticity is $K = 0.5$.}
\end{figure}

Assuming the power law for the energy dependences of the parameters 
$a_{1}$ and $a_{2}$, we obtain $(\sqrt{s} \; \mbox{in GeV})$ 
\begin{equation}    
\begin{array}{l}
     a_{1}= 0.915 (\sqrt{s})^{0.206} \\ 
     a_{2}= 0.289 (\sqrt{s})^{0.501}  
\end{array}
\end{equation}
which are shown in Fig.~10. \\

\section{Discussions}

\noindent
(i) {\it Energy dependence of the multiplicity}

Fig.~11 shows the energy dependence of the charged multiplicity, eq.(9),
by the present formulation, together with experimental data.\cite{alner87}
It describes the data well at high energies of $\sqrt{s} > 50 $ GeV.
The data at lower energies are described better by the curve of
the assumed case of the inelasticity $K =0.5$. (see the text below
for the value of the inelasticity at low energies) 

\begin{figure}[h]
\includegraphics[width=65mm]{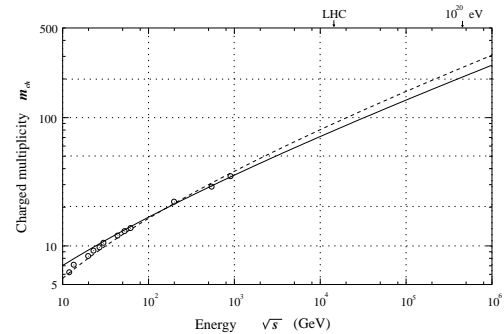}
\caption{Energy dependence of the charged multiplicity $m_{ch}$,
defined by eq.(9).
The solid line is  for the values of the parameters in eq.(13).
Experimental data are compiled by UA5 Collaboration.[4]
The chain line is for the assumed case of eq.(12), 
for which the inelasticity remains constant $(\sim 0.5)$.}
\end{figure}

\begin{figure}[h]
\includegraphics[width=65mm]{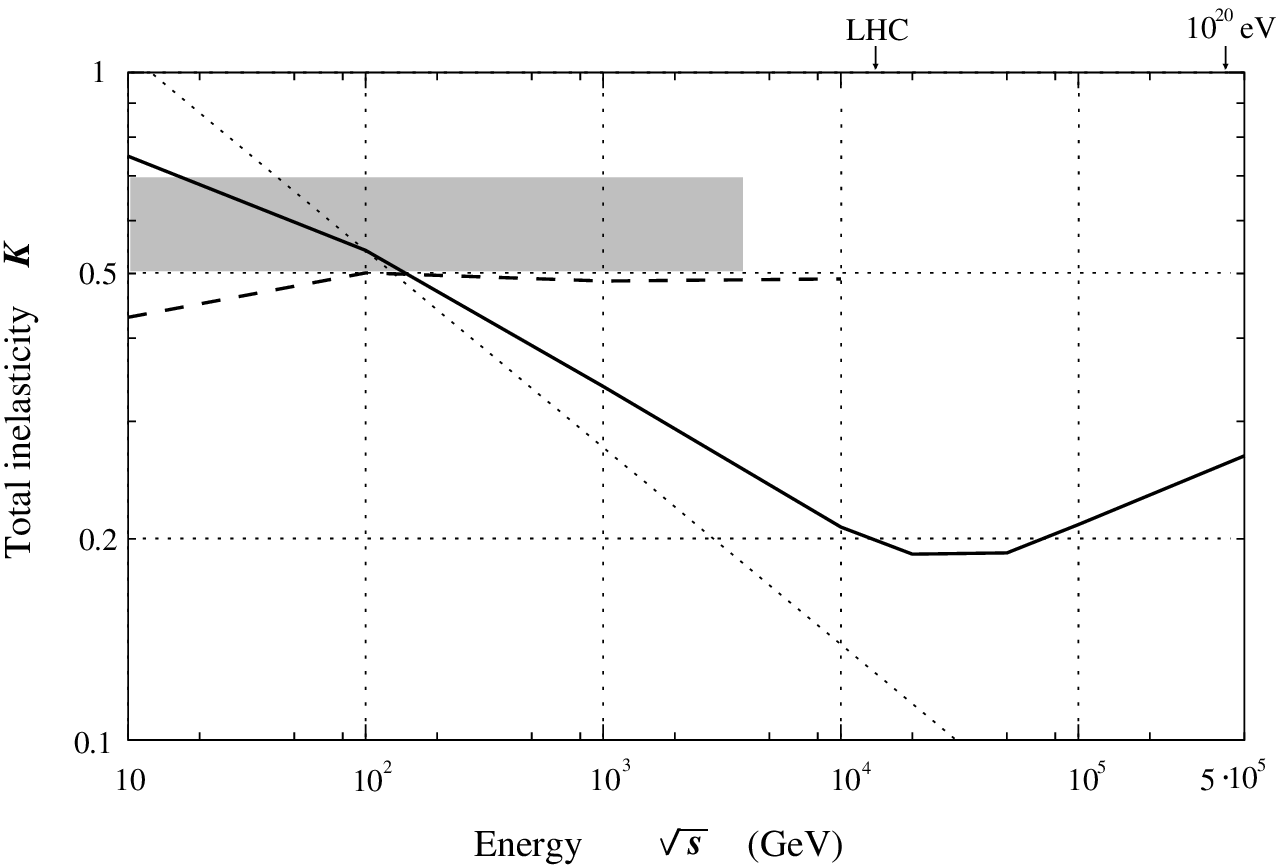}
\caption{Energy dependence of the total inelasticity, defined by eq.(10),
by the present formulation.~(the solid line)  The inelasticity decreases
with the energy in the energy region $\sqrt{s} < 2 \times 10^{4}$ GeV,
and then increases.  \\ 
\hspace*{2mm}The shaded area is the predictions by some models
of multiple particle production.[13]
Note that the predicted quantities
are not exact inelasticities but $(1-\eta')$'s where the parameter $\eta'$ is
the energy fraction of the highest energy baryon among the produced 
particles. (``Produced particles'' here include the surviving particle,
which is different from the definition in the present paper.) \\
\hspace*{2mm} The dotted line is $a_{1}/a_{2} \propto (\sqrt{s})^{-0.295}$,
which is steeper than the inelasticity due to the energy dependence
of the parameters $p_{1}$ and $p_{2}$ in the energy distribution.
The chain line is for the assumed case of eq.(12), for which the
inelasticity remains almost constant $(\sim 0.5)$.} 
\end{figure}

\bigskip\noindent
(ii) {\it Energy dependence of the inelasticity}

Fig.~12 shows the energy dependence of the total inelasticity, eq.(10),
by the present formulation. The inelasticity decreases with the energy
in the energy region $\sqrt{s} < 2 \times 10^{4}$ GeV, due to the
rapid increase of the parameter $a_{2}$, compared with the parameter $a_{1}$. 
(The inelasticity is proportional to $a_{1}/a_{2}$ approximately.)
After that it increases,
due to the exchange of the dominant parameter in the energy distribution
from $p_{1}$ to $p_{2}$ ($p_{1} < p_{2}$) through rapid increase 
of the parameter $r$.
(Experimental data of the inelasticity is $K \simeq 0.5$ at low energies
around $\sqrt{s} = 10$ GeV by bubble chamber experiments.)

\bigskip\noindent
(iii) {\it Pseudo-rapidity density distribution at LHC energy}
 
Fig.~13 shows the pseudo-rapidity density distribution
of the produced particles at $\sqrt{s}=14$ TeV 
(LHC energy) by the present formulation.  
Compared with the predictions by other models,
the shrinkage of the forward region is distinguished, which results 
in a small inelasticity of the present formulation.  
Predictions by the models are distributed widely, and even a single
data of the pseudo-rapidity density at $\eta^{*}=0$ can discriminate
some models as improbable if they cannot modify their predictions.\footnote{%
It is not self-evident that the model, tuned at LHC energy, reproduces
the low-energy data which are referred to tune it previously.}

According to CMS Collaboration the pseudo-rapidity density at $\eta^{*}=0$
at $\sqrt{s}=7$ TeV is $dN_{ch}/d\eta^{*} = 5.78 \pm 0.01\mbox{(stat)} 
\pm 0.23\mbox{(syst)}$.\cite{khachatryan10} It indicates that 
the increase of the parameter $a_{1}$ is more rapid than the one of
eq.(13) at $\sqrt{s} = 7$ TeV, which will be discussed elsewhere.  

The distribution of the surviving particle is shown together in Fig.13.
We assume that the charge exchange probability of the incident proton
into neutrons is 0.5, details of which are described in Ref.\cite{ohsawa10}.

\begin{figure}[h]
\includegraphics[width=65mm]{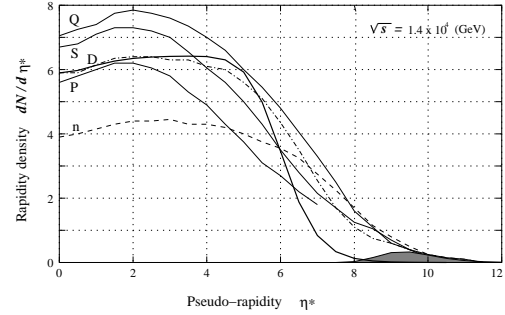}
\caption{Pseudo-rapidity density distribution of produced particles 
(the solid line) and the surviving particle (the shaded area) 
at $\sqrt{s}=14$ TeV (LHC energy).  The thin lines are the predictions
of the pseudo-rapidity density by some models.[14,15] \\
Q : QGSJET01, \ S : SIBYLL2.1, \ D : DPMJET2.55, \\
P : PYTHIA, \ n : neXus2.1}
\end{figure}

\bigskip\noindent
(iv) {\it Speculations related to the inelasticity to decrease and to increase}

According to the present formulation the inelasticity
is decreasing in the energy region $\sqrt{s} < 2 \times 10^{4}$ GeV
($E_{0} < 2 \times 10^{17}$ eV) and then increasing in $\sqrt{s} > 5 \times 
10^{4}$ GeV ($E_{0} > 10^{18}$ eV).
This structure of the inelasticity is caused by the rapid increase 
of the parameter $a_{2}$, compared with
that of $a_{1}$, which means that particle production is suppressed
strongly in the forward region.
  
The inelasticity is related to the attenuation mean free path
of cosmic rays, which
is a dominant factor to govern the cosmic-ray propagation in the 
atmosphere.  Small (Large) inelasticity
makes the attenuation mean free path long (short).
Consequently the development, rise and fall, of the air showers 
becomes slow (rapid), and
the air shower size at the maximum development is small (large) since
the total track length of the air shower particles is conserved.
Hence there is a possibility that following problems may be cleared
by the decreasing and increasing inelasticity. \\
(1) Intensity of the primary cosmic rays. \\ \indent
There is a discrepancy between the primary cosmic-ray intensities 
by balloon experiments of direct observation 
and by air shower experiments of indirect observation,
the former being lower than the latter by a factor 2.\cite{horandel08} \\ 
(2) $<X_{max}>$ and $RMS(X_{max})$ by Auger Collaboration \\ \indent
According to Pierre Auger Collaboration to observe highest energy air showers,
both the depth of shower maximum, $<X_{max}>$, and the dispersion 
of the maximum depth, $RMS(X_{max})$, are reaching the expected lines of
the iron primaries from those of the proton primaries
in the region $E_{0}= 10^{18} - 4 \times 10^{19}$ eV.\cite{abraham10}

Detailed and  quantitative discussions will be made elsewhere.

\end{document}